# Nonlinear Valley and Spin Valves in Bilayer Graphene

Xin Liao[1†], Xing-Yu Liu[1†], An-Qi Wang[1], Qing Yin[1], Tong-Yang Zhao[1], Zhi-Min Liao[1,2*]

[1]State Key Laboratory for Mesoscopic Physics and Frontiers Science Center for Nano-optoelectronics, School of Physics, Peking University, Beijing 100871, China.

[2]Hefei National Laboratory, Hefei 230088, China.

†These authors contributed equally to this work.

*Corresponding author. Email: liaozm@pku.edu.cn

**ABSTRACT:** Nonlinear transport plays a vital role in probing the quantum geometry of Bloch electrons, valley chirality, and carrier scattering mechanisms. The nonlinear Hall effect, characterized by a nonlinear scaling of Hall voltage with longitudinal current, has been explored to reveal the Berry curvature and quantum metric related physics. In this work, we extend the study of nonlinear transport to spin and valley degrees of freedom. Using bilayer graphene devices with $Fe_3GeTe_2$ contacts, we observe a second-order nonlinear spin current exhibiting spin valve-like behaviors. By tracking magnetic moment precession under an in-plane magnetic field, we identify a significantly enhanced critical magnetic field required for in-plane rotation, suggesting out-of-plane valley polarization induced by ferromagnetic proximity. These findings offer deep insights into the interplay of valley and spin in second-order nonlinear transport, opening avenues for promising device applications.

## I. INTRODUCTION

Valleytronics has attracted enormous interest for next-generation electronics, in which the valley degree of freedom is used to encode, process and store information [1-3]. With both valley and spin degeneracies, graphene systems are promising candidates for valley- and spin-based quantum devices, due to high carrier mobility, long spin relaxation time, unique valley properties, and excellent gate tunability [4-9]. Experimentally, valley currents can be generated through circularly polarized light [10-12], magnetic proximity [13-16], or the valley Hall effect [17-19], where the valley current is linearly proportional to the applied electric field. Recently, nonlinear transport has been found to correlate with quantum geometry [20-22] and valley-contrasting scattering [23, 24], leading to potential applications in frequency doubling and rectification [25-27]. However, the exploration of valley and spin transport in the nonlinear regime, alongside thermoelectric studies [28], remains uncharted.

Bilayer graphene (BLG) presents the potential for a tunable bandgap under an applied electric displacement field [29-31], hosting substantial Berry curvature and a significant valley Hall effect [32]. It allows a nonzero second-harmonic generation once the inversion symmetry is broken. Recent studies have revealed that the valley-contrasting scattering, attributed to valley chirality, underpins the second-order nonlinear transport in twisted bilayer graphene [25] and graphene-hBN superlattices [26]. In contrast to the Berry curvature dipole-induced nonlinear Hall effect observed in $T_d$-WTe$_2$ [33-35] and TaIrTe$_4$ [36], skew scattering generates second-order nonlinear currents in both transverse and longitudinal directions [25, 26], broadening the range of materials capable of nonlinear response and potentially yielding more pronounced signals. Crucially, valley-contrasting scattering presents a unique opportunity for manipulating the valley degree of freedom in the nonlinear regime [23, 24].

To investigate valley and spin nonlinear transport properties, we fabricate BLG devices with Fe$_3$GeTe$_2$ (FGT) contacts of different thickness (see Appendix A). Upon applying an AC current, a longitudinal second-order voltage is observed, exhibiting high- and low-voltage states under parallel and antiparallel magnetizations of FGTs. It acts like a nonlocal spin valve signal [7-9] but scales quadratically with the applied current. The observed nonlinear signal depends on the direction of the magnetic field owing to the strong perpendicular magnetic anisotropy of FGT. Furthermore, spin precession is revealed under an in-plane magnetic field, but a notably high critical magnetic field (about 800 mT) is required to flop the spins to the in-plane direction. This indicates potential out-of-plane orbital magnetization associated with valley polarization induced by ferromagnetic proximity.

## II. METHODS

*Device Fabrication.* The bilayer graphene devices with two $Fe_3GeTe_2$ (FGT) contacts were fabricated through dry transfer process. Prior to the transfer process, all van der Waals materials, including hBN, FGT, BLG, and graphite flakes, were mechanically exfoliated onto separate $Si/SiO_2$ (285 nm) substrates. Step I: (1) A suitable hBN flake and a thin graphite layer were successively picked up and transferred onto a $Si/SiO_2$ substrate using a polymer-based dry transfer technique. (2) Ti (2 nm)/Au (10 nm) electrodes were deposited onto the bottom hBN flake through electron-beam lithography and evaporation. (3) Surrounding Ti (5 nm)/Au (45 nm) electrodes were fabricated in the same manner on the $Si/SiO_2$ substrate to establish connections with the electrodes on hBN. Step II: (1) Thin graphite, hBN, two FGTs, and BLG were sequentially picked up and transferred onto the bottom electrodes using the same dry transfer technique. (2) The device underwent a chloroform soak to eliminate any residual polymer, followed by a thorough acetone rinse. The entire process of mechanical exfoliation and transfer was conducted within a glove box environment with an argon atmosphere.

*Nonlinear Electrical Transport Measurements.* Electrical transport measurements were conducted within a commercial Oxford cryostat. The sample stage can tilt from 0° to 360°, so the direction of magnetic field can be changed relatively. The top and bottom gate voltages were applied via two Keithley 2400 SourceMeters. Standard lock-in techniques were used to measure the first- and second-harmonic transport signals utilizing the Stanford Research SR830 and SR865A instruments. An AC current $I_\omega$ with the various frequencies (17.777 Hz, 47.77 Hz, 77.77 Hz, and 177.77 Hz) was applied to the devices. And the phases of lock-in amplifiers were set to 0° and -90° for the first- and second-harmonic voltage measurements, respectively. Specifically, the second-harmonic voltage is $V_{2\omega} \propto (I\sin\omega t)^2 = I^2[1 + \sin\left(2\omega t - \frac{\pi}{2}\right)]/2$, directly suggesting the -90° phase shift of the second-harmonic signals.

## III. RESULTS AND DISCUSSION

The schematic illustration of a BLG device is depicted in Fig. 1(a). Two FGT contacts are transferred on the BLG flake, and a dual gate configuration is utilized to independently control the carrier density and displacement field. By applying a vertical displacement field, the inversion symmetry can be effectively broken, thereby tuning the second-harmonic generation in BLG. Figure 1(a) illustrates a four-terminal longitudinal measurement with standard lock-in techniques. An AC current $I_\omega$ with driving frequency $\omega$ is applied, and the two FGT contacts are used to measure the first- and second-order longitudinal voltages.

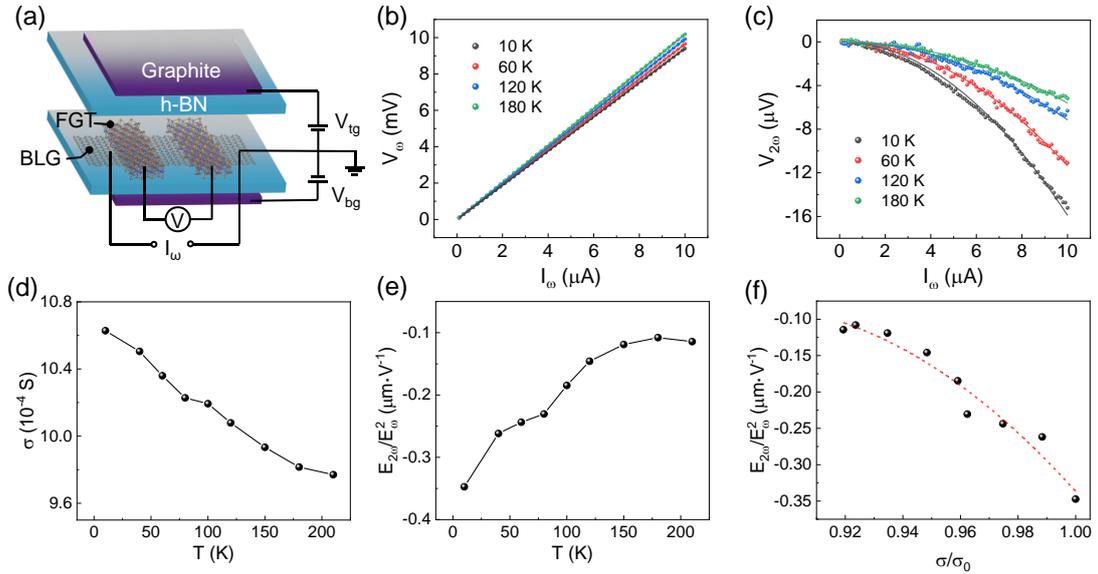

FIG. 1. (a) Schematic of the dual-gated BLG device. Two FGT contacts are placed on the BLG flake as the voltage probes. (b, c) First- and second-harmonic voltages, $V_\omega$ and $V_{2\omega}$, plotted against the AC current amplitude $I_\omega$ at $V_{tg} = -3.76\,\text{V}, V_{bg} = 0\,\text{V}$

at various temperatures, respectively. Solid lines indicate linear and quadratic fits. (d, e) Temperature dependence of $\sigma$ and $E_{2\omega}/E_\omega^2$, respectively, with solid lines as visual guides. (f) $E_{2\omega}/E_\omega^2$ as a function of $\sigma/\sigma_0$, where the dashed line indicates a parabolic fit. All data are measured in Device A.

We first conduct measurements across temperatures ranging from 10 to 210 K at $V_{tg} = -3.76$ V under parallel magnetizations of FGTs. The first-harmonic voltage ($V_\omega$) exhibits a linear dependence on $I_\omega$ [Fig.1(b)], signifying the good ohmic contacts. In contrast, the second-harmonic voltage ($V_{2\omega}$) shows a quadratic dependence on $I_\omega$ [Fig. 1(c)]. Both the conductivity ($\sigma$) and $V_{2\omega}$ decreases with increasing temperature [Fig. 1(d)], acting as metallic behaviors due to phonon scattering [37]. The nonlinear coefficient, $\frac{E_{2\omega}}{E_\omega^2} = \frac{V_{2\omega}L}{V_\omega^2}$, where $L$ is the device's channel length, decreases with increasing scattering strength, as depicted in Fig.1(e), aligning with findings reported in graphene superlattices [25, 26]. By applying scaling law of second-harmonic transport [38], expressed as

$$\frac{E_{2\omega}}{E_\omega^2} = A_0 + A_1\frac{\sigma}{\sigma_0} + A_2\left(\frac{\sigma}{\sigma_0}\right)^2$$

where $\sigma_0$ denotes the zero-temperature conductivity, we can distinguish the different contributions, like intrinsic, side-jump, and skew-scattering contributions. The measured conductivity at 10 K is used as $\sigma_0$ for scaling analysis to mitigate weak localization effects at lower temperatures [39]. The fitting results are plotted in Fig. 1(f), and the extracted three fitting parameters, $A_0$, $A_1$ and $A_2$, have a special relation: $A_0 : A_1 : A_2 \approx 1 : -2 : 1$ [1, 2, 26]. Besides, it is well consistent at different gate voltages (see Appendix B). This feature strongly points to the dominant contribution of skew scattering due to dynamic disorders (like phonons), aligning with the preserved three-fold rotation symmetry in BLG (detailed discussions in Appendix F).

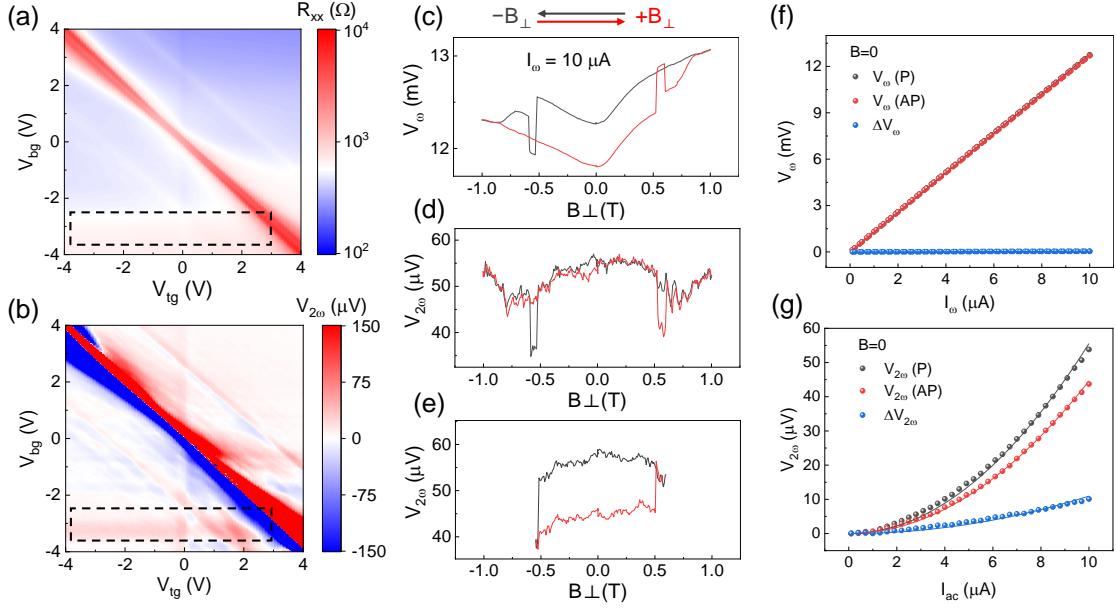

FIG. 2. (a, b) Top and bottom gate mapping of longitudinal resistance $R_{xx}$ and second-harmonic voltage $V_{2\omega}$, respectively, measured at 1.6 K. Dashed rectangles outline the charge neutral points of the proximity-modified BLG underneath FGT contacts. (c, d) First- and second-harmonic voltages, $V_\omega$ and $V_{2\omega}$, as a function of the out-of-plane magnetic field with an applied current $I_\omega = 10$ μA at $V_{tg} = V_{bg} = -3$ V. (e) The minor hysteresis loop of $V_{2\omega}$ when sweeping the magnetic field to selectively reverse the magnetization of one FGT contact. (f, g) Current dependence of $V_\omega$ and $V_{2\omega}$ in parallel (P) and antiparallel (AP) magnetizations of two FGT contacts, and $\Delta V_\omega$ (or $\Delta V_{2\omega}$) is the variation between the two states. The data in (f, g) are symmetrized to exclude Hall mixing (see Appendix C), and solid lines represent linear and quadratic fits to the data. All data are measured in Device A.

We then perform systematic measurements at 1.6 K. By tuning both the top and bottom gate voltages, $V_{tg}$ and $V_{bg}$, the maps of the longitudinal resistance $R_{xx}$ and second-harmonic voltage $V_{2\omega}$ are shown in Figs. 2(a) and 2(b). The sharp peaks of $R_{xx}$ near charge neutral points (CNPs) indicate the high mobility of BLG, approximately $10^4$ cm$^2$ V$^{-1}$ s$^{-1}$. Along the CNP line in Fig. 2(a), the peak value of $R_{xx}$ increases substantially as the bandgap opens due to the applied vertical displacement field, corresponding to the effective inversion symmetry breaking in BLG. However, unexpected peaks of $R_{xx}$ are observed around $V_{bg} = -3.5$ V, outlined by the dashed rectangles in Fig. 2(a). The invariance of $R_{xx}$ with respect to varying $V_{tg}$ is attributed to the electrostatic shielding of FGT contacts, indicating a finite *n*-doping of about 4.5 × 10^{12} cm^{-2} in the proximity-modified region of BLG beneath the FGT contacts. Furthermore, the longitudinal second-harmonic voltages $V_{2\omega}$ exhibits a significant increase near the peaks of $R_{xx}$, with clear sign reversals at CNPs as the type of carrier transitions. Correspondingly, significant $V_{2\omega}$ is also observed around $V_{bg} = -3.5$ V, suggesting CNPs of the proximity-modified BLG.

To investigate the spin-related transport properties, we sweep the out-of-plane magnetic field to switch the magnetizations of the two FGT contacts. An antisymmetric hysteresis loop of $V_{\omega}$ is observed in the raw data [Fig. 2(c)], indicating a mixture of Hall voltages owing to the unavoidable misalignment during device fabrication. Two different coercive fields, 0.52 T and 0.60 T, correspond to the different thicknesses of two FGT flakes, about 7 nm and 8 nm, respectively. In contrast, near the CNPs of proximity-modified BLG at $V_{bg} = V_{tg} = -3$ V, $V_{2\omega}$ exhibits symmetric variations at the coercive fields with high- and low-voltage states [Fig. 2(d)], excluding possible Hall mixing. As sweeping the magnetic field over a narrower range and selectively reversing the magnetization of one FGT contact, a discernible hysteresis loop of $V_{2\omega}$ is observed [Fig. 2(e)], indicating the stability of the second-harmonic signals in parallel and antiparallel magnetizations of FGT contacts. Similar behavior was also observed in Device B (see Appendix D), resembling nonlocal spin valve signals, where a spin current is injected from one ferromagnet and detected by another. However, the observed $V_{2\omega}$ here varies quadratically with the injected current $I_{\omega}$ [Fig. 2(g)], and the measurements are performed in a four-terminal configuration. Thereby, we successfully expand the spin current into the second-order nonlinear regime, showcasing the novelty and potential of spintronics for frequency-doubling and rectification applications.

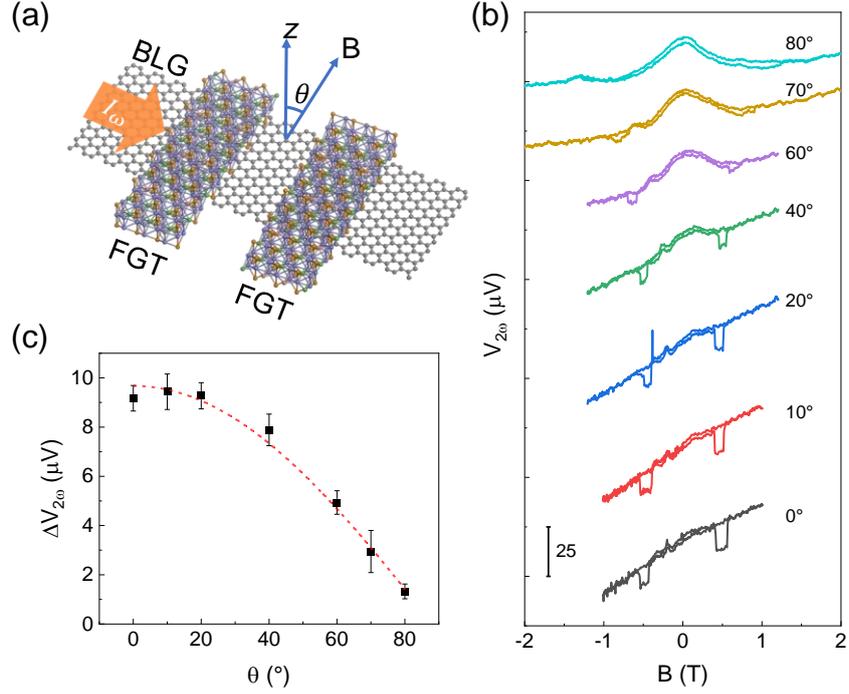

FIG. 3. (a) Electrical measurement configuration, where $\theta$ represents the angle between the external magnetic field and the out-of-plane $z$-axis. At $\theta = 0°$, the magnetic field is perpendicular to the BLG plane. (b) Second-harmonic voltages $V_{2\omega}$ plotted against magnetic fields along different orientations. Curves are offset for clarity. (c) Extracted $\Delta V_{2\omega}$ as a function of $\theta$. The dashed line shows a $\cos\theta$ fit to $\Delta V_{2\omega}$. All data are acquired with an excitation current $I_\omega = 30\ \mu\text{A}$, at $V_{bg} = -2.6\ \text{V}, V_{tg} = 0\ \text{V}$ and temperature of 1.6 K in Device B.

For further exploration, we measure second-harmonic voltages under various orientations of the magnetic field. Figure 3 illustrates the angular-dependent measurement results, where the tilted angle $\theta$ is defined as the angle between the magnetic field and the normal line of BLG. As $\theta$ increases, the coercive fields of FGT contacts increase, particularly at larger angles ($\theta > 60°$) [Fig. 3(b)]. This corresponds to the robust perpendicular magnetic anisotropy of FGT [40-42]. Conversely, under a titled nonzero magnetic field, spins in the BLG undergo precession and relaxion, eventually aligning with the magnetic field direction. As shown in Fig. 3(c), $\Delta V_{2\omega}$ decreases as the direction of magnetic field varies from the out-of-plane direction ($\theta =$ 0°) to nearly in-plane direction ($\theta = 80°$), indicating the out-of-plane direction of spins under zero magnetic field. It is well consistent with the nonlocal spin valves in Fe$_3$GeTe$_2$/graphene [43] and Fe$_3$GaTe$_2$/graphene [44] heterostructures.

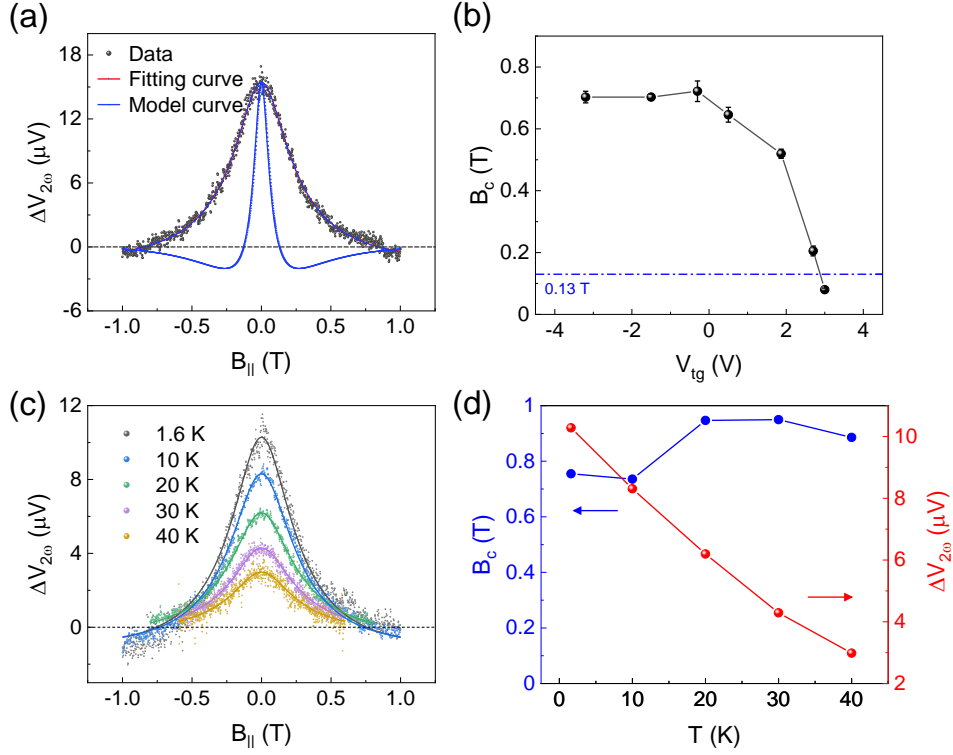

FIG. 4. (a) Hanle spin precession measured as a function of in-plane magnetic field $B_{\parallel}$ at $V_{tg} = V_{bg} = -3.2$ V and a temperature of 1.6 K. The red line represents the best fits based on classical Hanle equation. The blue line represents a typical Hanle spin precession with spin relaxation time $\tau_s = 100$ ps and the diffusion constant $D = 0.03$ m$^2$/s. (b) The critical field $B_c$ extracted from Hanle curves measured at different $V_{tg}$ with $V_{bg} = -3.2$ V. (c) Hanle curves at different temperatures at $V_{tg} = V_{bg} = -3$ V. The solid lines are Hanle curve fitting. (d) Temperature dependence of $B_c$ and the amplitude of $\Delta V_{2\omega}$ extracted from the fitted curve in (c). All data are measured in Device A.

Finally, we perform spin precession measurements under an in-plane magnetic field to provide deeper insights. Figure 4(a) illustrates the measured $\Delta V_{2\omega}$ as a function of in-plane magnetic field $B_{\parallel}$. Before applying the in-plane magnetic field, two FGT contacts are prepared in four magnetization states, then obtain $\Delta V_{2\omega}$ from the variation between parallel and anti-parallel states (see Appendix C). Referring to the spin precession model for Hanle effects [45], the applied $B_{\parallel}$ exerts a spin torque $\left(\frac{g\mu_B}{\hbar}\right)\vec{S}\times\vec{B}_{\parallel}$ and alters the spin direction at the Larmor frequency $\omega_L=\frac{g\mu_B B_{\parallel}}{\hbar}$, where $g$ is the g-factor of carriers in the channel, $\mu_B$ is the Bohr magneton, $\hbar$ is the reduced Planck constant and $\vec{S}$ is the spin of carriers. Usually, for materials like BLG with negligible spin-orbit coupling, the g-factor is identified as $g{\sim}2$. Upon reaching a critical value $B_c$, the magnetic field $B_{\parallel}$ could induce an average rotation angle of 90° on the spins during the diffusion between the two FGT contacts, causing the measured spin precession signal $\Delta V_{2\omega}$ to approach zero.

We estimate the typical spin relaxation time $\tau_s=100$ ps and the diffusion constant $D=0.03$ m$^2$/s, calculated using the Einstein relation $\sigma=e^2 N_\varepsilon D$, where $N_\varepsilon=g_v\frac{m_e^*}{\pi\hbar^2}=\frac{2m_e^*}{\pi\hbar^2}$ is roughly regarded as the density of state of a two-dimensional free electron gas [7, 8]. However, this estimated Hanle curve significantly deviates from the experimental data [Fig. 4(a)]. The critical field $B_c$ appears strongly larger than the estimated value ~0.13 T and those reported in previous studies [7, 8], typically around 100 mT or smaller. It indicates a much smaller spin torque generated by $B_{\parallel}$ in our devices. We vary $V_{tg}$ to change the charge density of the BLG channel, and the extracted $B_c$ is presented in Fig. 4b. It is found that $B_c$ decreases with increasing $V_{tg}$. Furthermore, the observed $B_c$ near the CNP is close to the estimated value, indicating that the conventional spin precession mechanism becomes dominant. Additionally, the temperature dependence of the Hanle effect is depicted in Figs. 4(c) and 4(d). Although the amplitude of $\Delta V_{2\omega}$ decreases with increasing temperature, the critical field $B_c$ remains nearly unchanged below 40 K.

The large deviation of our experimental data from the estimated model suggests underlying mechanisms to be explored. The critical magnetic field $B_c$ depends on three key transport parameters: the magnitude of spin torque (g-factor), the spin relaxation time $\tau_s$ and the diffusion constant $D$. However, the small variations of $\tau_s$ and $D$ cannot cause a six- to eight-fold change of $B_c$ in the similar graphene systems, as reported in previous experiments [8, 46]. And thus, the reduction of torque exerted by the applied in-plane magnetic field $B_{\parallel}$ could be the dominant factor, implying a reduction of the g-factor and potential orbital contributions. It is essential to underscore that the orbital motions of electrons in a two-dimensional plane can only generate an out-of-plane orbital magnetic moment [1]. This orbital magnetic moment is hard to rotate under an in-plane magnetic field, thereby accounting for the decrease of the generated torque and the concurrent increase in $B_c$.

Here, we consider the proximity-modified BLG underneath the FGT contacts as the second-harmonic spin or valley current source (detailed discussions in Appendixes E and F). The difference of $V_{2\omega}$ between parallel and antiparallel magnetizations of FGTs solely emerge near the CNPs of the proximity-modified BLG [Fig. 2(e)]. Consequently, only the low-energy electrons (holes) near the bottom (top) of the K and K' valleys are concerned. This discussion addresses two primary consequences of ferromagnetic proximity, containing the spin Zeeman splitting [47] and the valley Zeeman effect [13-16]. The former results in a band splitting of spins in opposite directions through out-of-plane magnetic proximity, thereby generating spin polarization. The latter is anticipated to induce opposing energy shifts in K and K' valleys, fostering a valley polarization through perpendicular magnetic proximity. The orbital magnetic moment, denoted as $m_{orb}$, exhibits opposite signs in disparate valleys and converges near the CNP, exceeding 30 times the Bohr magneton according to prior studies [1]. Consequently, the second-harmonic currents emanating from the ferromagnetic proximity-modified BLG are characterized not only by spin polarization but also by carrying nonzero out-of-plane orbital magnetic moments. This dual nature contributes significantly to mitigating the torque exerted by an in-plane magnetic field.

## IV. CONCLUSION

In summary, we have observed second-harmonic voltages with spin and valley polarization in BLG utilizing two FGT contacts. Our scaling analysis suggests the skew scattering origin of the longitudinal second-harmonic currents. It shows high- and low-voltage states under parallel and antiparallel magnetizations of two FGT contacts, acting as a spin valve. But differently, the voltage variation between the two states is frequency-doubled and scales quadratically with the applied AC current. Additionally, under an in-plane magnetic field, the coherent spin precession is observed but with a much larger critical magnetic field—defined as the magnetic field where the average angle of spin precession reaches about 90°. Our analysis proposes that the reduction of the spin torque exerted by the in-plane magnetic field originates from potential orbital contributions associated with valley polarization induced by the ferromagnetic proximity. Our work extends spin transport in graphene systems into the second-harmonic regime for the first time, paving the way for novel applications in spintronics and valleytronics, including rectification and frequency-doubling in gigahertz or terahertz frequencies.

**ACKNOWLEDGEMENTS**

This work was supported by the National Natural Science Foundation of China (Grant Nos. 62425401 and 62321004), and Innovation Program for Quantum Science and Technology (Grant No. 2021ZD0302403).

**APPENDIX A: OPTICAL IMAGE AND CHARACTERIZATION OF BILAYER GRAPHENE DEVICES**

The optical images of Device A and B are shown in Figs. 5(a) and 5(b), where the BLG flake and two FGT electrodes are clearly marked. We measure the magnetoresistance of Device A under an out-of-plane magnetic field. The results are shown in Fig. 5(c), revealing clear Shubnikov-de Haas (SdH) oscillations, indicating the high quality of Device A. The extracted SdH fan diagram in Fig. 5(d) shows an intercept near zero, consistent with the $2\pi$ Berry phase of bilayer graphene [48, 49].

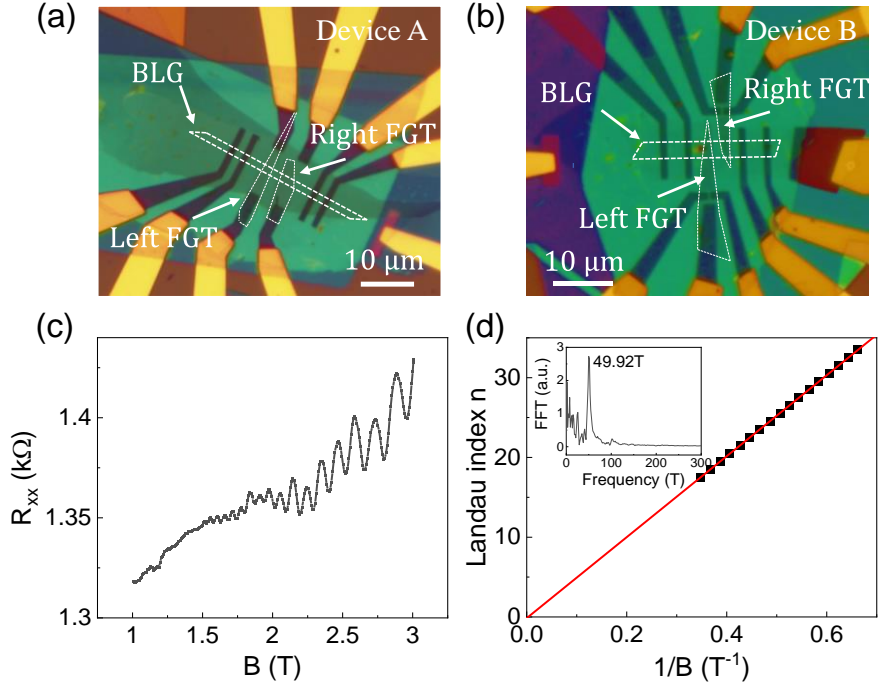

FIG. 5. (a, b) Optical images of the dual-gate BLG Device A and B. The BLG and FGT flakes are marked by dashed lines. (c) Magnetoresistance of Device A under an out-of-plane magnetic field with clear SdH oscillations measured at $V_{tg} = V_{bg} = -3$ V at 1.6 K. (d) The Landau level index plot $n$ vs $1/B_\perp$ extracted from (c). The red line is the linear fit with an intercept close to 0. Inset of (d): fast Fourier transform (FFT) of $R_{xx}$ against $1/B_\perp$ with only one dominating frequency of $B_F = 49.92$ T.

## APPENDIX B: ADDITIONAL SCALING ANALYSIS OF DEVICE A

Figure 6 presents the second-harmonic transport results measured at various temperatures. The longitudinal resistance $R_{xx}$ and the second-harmonic voltage $V_{2\omega}$ are measured, as shown in Figs. 6(a-d). We calculate the nonlinear coefficient, $E_{2\omega}/E_\omega^2 = V_{2\omega}L/V_\omega^2$, and the conductivity ratio $(\sigma/\sigma_0)$, where $L$ is the BLG channel length approximately 2 μm (the closest distance between two FGTs). Gate voltages of $V_{tg} = -2.56$ V and 2.6 V were selected as supplementary, as the BLG at both gate voltages exhibits clear metallic behavior, effectively excluding thermal activation effects at high temperatures. Figures. 6(e, f) depict the fitting results based on the scaling relation in the main text. The fitting parameters, $A_0$, $A_1$ and $A_2$, have the same special relation, $A_0 : A_1 : A_2 \approx 1 : -2 : 1$, indicating that skew scattering is the dominant mechanism for the longitudinal second-harmonic voltages, effectively excluding other possible origins.

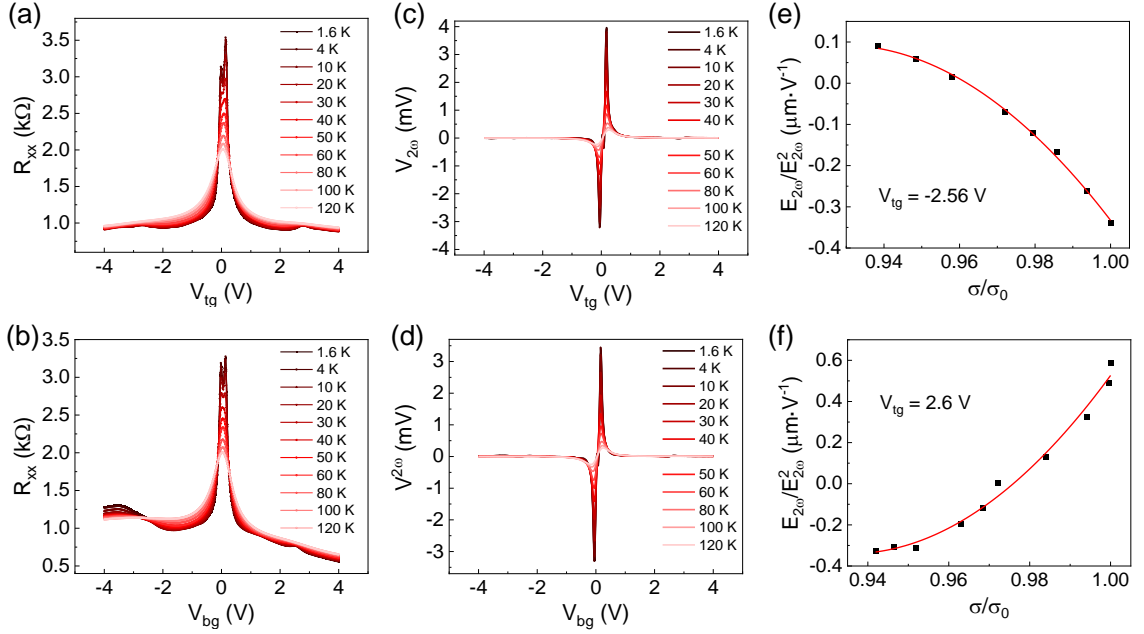

FIG. 6. (a, b) Gate voltages, $V_{tg}$ and $V_{bg}$, dependence of the longitudinal resistance $R_{xx}$ at different temperatures. (c, d) Gate voltages, $V_{tg}$ and $V_{bg}$, dependence of the second-order voltage $V_{2\omega}$ measured with an applied current $I_\omega = 5\ \mu A$ at different temperatures. (e, f) The nonlinear coefficient $E_{2\omega}/E_\omega^2$ as a function of $\sigma/\sigma_0$ at $V_{tg} = -2.56\ \text{V}$ and $2.6\ \text{V}$.

## APPENDIX C: SYMMETRIZATION METHOD

In experiments, two FGTs were used as electrodes. However, FGTs exhibits obvious anomalous Hall effect, mixed with longitudinal signals. The first- and second-order longitudinal voltages can be separated by using a symmetrization method. By sweeping magnetic field, four magnetization states of two FGTs may arise: ↑↑, ↑↓, ↓↑ and ↓↓ (red and blue arrows indicate the magnetization of the two FGTs). To excluding Hall mixing, the data were symmetrized as follows:

$$V_{n\omega}(P) = \frac{V_{n\omega}^{\uparrow\uparrow} + V_{n\omega}^{\downarrow\downarrow}}{2},$$

$$V_{n\omega}(AP) = \frac{V_{n\omega}^{\uparrow\downarrow} + V_{n\omega}^{\downarrow\uparrow}}{2},$$

$$\Delta V_{n\omega} = V_{n\omega}(P) - V_{n\omega}(AP),$$

where $n = 1$ or $2$ denotes the harmonic order index.

## APPENDIX D: ADDITIONAL DATA OF DEVICE B

Figure 7 shows additional measurements of Device B. The *n*-doping of the BLG beneath FGT flakes, along with the additional resistance peak at $V_{bg} = -2.6$ V, is clearly visible in Fig. 7(a). Similar to Device A, antisymmetric and symmetric variations of $V_{\omega}$ and $V_{2\omega}$ with the out-of-plane magnetic field are observed, respectively, at $V_{bg} = -2.6$ V, $V_{tg} = 0$ V [Figs. 7(b) and 7(c)]. Additionally, a minor loop of $V_{2\omega}$ is obtained when sweeping the magnetic field to flip the magnetization of only one FGT contact [Fig. 7(d)]. As sweeping the back gate and top gate voltages, the $V_{2\omega}$ mappings for four magnetization states of the FGT electrodes under zero magnetic field are obtained. The mapping of $\Delta V_{2\omega}$, defined as the difference between parallel and antiparallel states, is shown in Fig. 7(e), with pronounce signals near $V_{bg} = -2.6$ V.

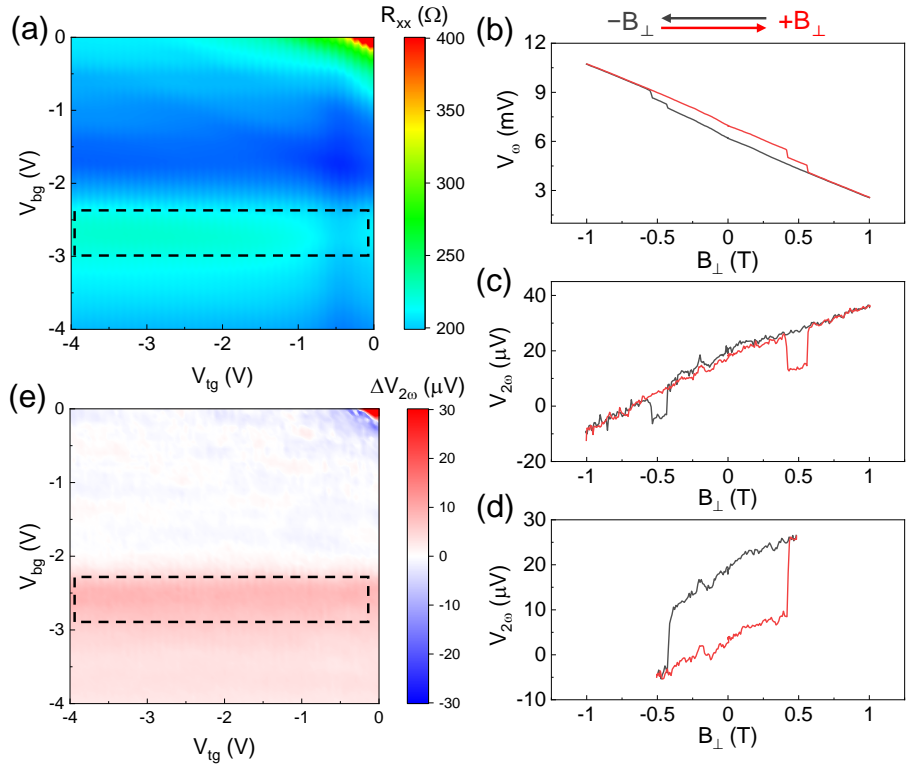

FIG. 7. (a) Gate mapping of longitudinal resistance $R_{xx}$. (b, c) First- and second-harmonic voltages, $V_{\omega}$ and $V_{2\omega}$, as a function of the out-of-plane magnetic field with an applied current $I_\omega = 30$ µA at $V_{bg} = -2.6$ V, $V_{tg} = 0$ V. (d) The minor hysteresis loop of $V_{2\omega}$ with the sweeping magnetic field of $\pm 0.5$ T. (e) Gate mapping of $\Delta V_{2\omega}$. All data are measured in Device B.

## APPENDIX E: SPIN PROCESSION MODEL

The spin orientation of carriers in the BLG channel has been confirmed to be out-of-plane, as evidenced by angular-dependent experiments (Fig. 3). Under an in-plane magnetic field $B_\parallel$, the spins can be rotated by the external torque $\left(\frac{g\mu_B}{\hbar}\right)\vec{S} \times \vec{B}_\parallel$, and relax over the transit time. Consequently, the detected spin signals is modulated by three key parameters: the magnitude of spin torque (g-factor), the spin relaxation time $\tau_s$, and the diffusion constant $D$. Besides, in the four-terminal measurement configuration, the spin attains a nonzero drift velocity $v = \mu E$, where $\mu$ is the mobility. Therefore, the quantitative interpretation of our Hanle results is based on the drift-diffusion model [45], with the spin-related voltage given by:

$$\Delta V_{2\omega} \propto \int_0^{+\infty} \frac{1}{\sqrt{4\pi Dt}} e^{-\left[\frac{(L-vt)^2}{4Dt}\right]} \cos(\omega_L t)\, e^{-\frac{t}{\tau_s}} dt$$

where $L$ is the channel length and $\omega_L$ is the Larmor frequency. While the experiment results fit this model well, the extracted $D$ is one order of magnitude larger than the estimation, and $\tau_s$ is one order of magnitude smaller, assuming the g-factor $g = 2$. As shown in Fig. 4(a), the estimated curve using typical values deviates significantly from the experimental results, suggesting that the spin torque exerted by $B_\parallel$ may be much smaller. It indicates the possible orbital contributions, perhaps related to valley polarization induced by the ferromagnetic proximity.

## APPENDIX F: ORIGINS OF SECOND-HARMONIC SIGNALS

### 1. Analysis of second-harmonic longitudinal resistance

The BLG under a nonzero vertical displacement field remains the $C_{3v}$ symmetry. When an in-plane electric field is applied, the generated second-harmonic currents have only in-plane components. Due to the $C_{3v}$ symmetry constraints, the second-harmonic conductivity tensor has one independent nonzero component. Under an electric field $\boldsymbol{E}^{\omega} = (E_x^{\omega}, E_y^{\omega})$, the generated second-harmonic current $\boldsymbol{j}^{2\omega} = (j_x^{2\omega}, j_y^{2\omega})$ is given by

$$\begin{pmatrix} j_x^{2\omega} \\ j_y^{2\omega} \end{pmatrix} = \begin{pmatrix} 0 & 0 & \sigma^{2\omega} \\ \sigma^{2\omega} & -\sigma^{2\omega} & 0 \end{pmatrix} \begin{pmatrix} (E_x^{\omega})^2 \\ (E_y^{\omega})^2 \\ 2E_x^{\omega}E_y^{\omega} \end{pmatrix}$$

where $x$ and $y$ correspond to the zigzag and armchair directions of BLG, and $\sigma^{2\omega}$ is the nonzero component of second-harmonic conductivity. Therefore, the second-harmonic currents along the longitudinal ($V_{\parallel}^{2\omega}$) and transverse ($V_{\perp}^{2\omega}$) directions are proportional. The ratio of $V_{\parallel}^{2\omega}$ and $V_{\perp}^{2\omega}$ is a sample-dependent constant determined by the relative orientation of the crystal axis, allowing the data of $V_{\parallel}^{2\omega}$ to be modeled through the processing method for $V_{\perp}^{2\omega}$.

According to the theoretical study, the different contributions to nonlinear Hall effect, including intrinsic, side-jump, and skew scattering contributions, can be efficiently extracted from the scaling law analysis. The fitting parameters are expressed as

$$A_0 = C_{int} + C_1^{sj} + C_{11}^{sk,1}$$

$$A_1 = C_{01}^{sk,1} - 2C_{11}^{sk,1} + C_0^{sj} - C_1^{sj}$$

$$A_2 = C^{sk,2}\sigma_0 + C_{00}^{sk,1} + C_{11}^{sk,1} - C_{01}^{sk,1}$$

where $C_{int}$ is the intrinsic contribution, $C_i^{sj}$ is the side-jump contribution, $C_{ij}^{sk,1}$ is the Gaussian skew scattering contribution, $C^{sk,2}$ is the non-Gaussian skew scattering contribution, and $i, j = 0,1$ are the static and dynamic disorders. Considering that the intrinsic ($C_{int}$) contribution only affects $A_0$, it is unlikely to dominate due to the comparable magnitudes of $A_0$, $A_1$, and $A_2$. Similarly, the non-Gaussian skew scattering ($C^{sk,2}$) contributes sorely to $A_2$, making it an unlikely dominant role. Additionally, the side-jump ($C_i^{sj}$) plays a secondary role as its absence in $A_2$. In contrast, the special ratio of $A_0 : A_1 : A_2 = 1 : -2 : 1$, strongly points to dynamic skew scattering ($C_{11}^{sk,1}$) as the dominant origin.

## 2. Excluding of the side effects

We systematically consider and exclude various side effects that could contribute to second-harmonic signals as follows: <u>Diode Effect:</u> An accidental diode at the contact may induce nonreciprocal transport and rectification. Two-probe DC measurements reveal clear linear I-V characteristics [Fig. 8(a)], indicating negligible diode contributions. <u>Capacitive Effect:</u> Spurious capacitive coupling, potentially arising at contacts or in the measurement circuits, can induce rectification. However, the obtained second-harmonic voltage $V_{2\omega}$ is nearly independent of the driving frequencies [Figs. 8(b-d)], ruling out capacitive effects. <u>Thermoelectric Effect:</u> Joule heating from the injected current may create a possible temperature gradient ($\Delta T$) in the presence of inhomogeneity, generating a thermoelectric voltage $V_{th} \propto \Delta T \propto I^2 R$, and thus leading to a second-harmonic signal. According to the Mott formula, the typical Seebeck coefficient is proportional to temperature [50]. However, the observed decrease in $V_{2\omega}$ with increasing temperature [Fig. 1(c)] contradicts thermoelectric behavior. Besides, the well-fitting scaling relation supports the skew scattering mechanism, further excluding this side effect.

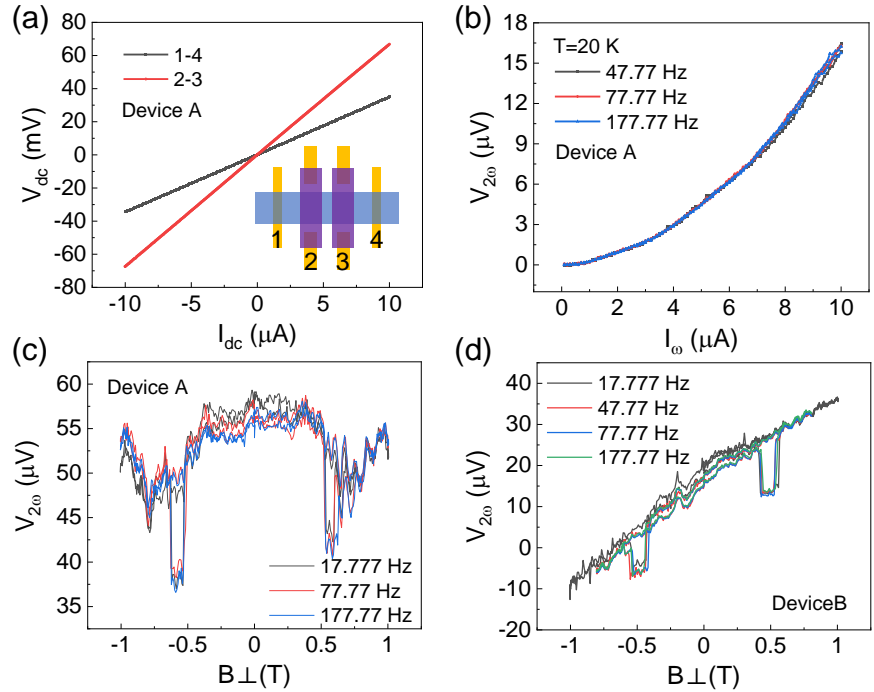

FIG. 8. (a) $I-V$ curves obtained from the two-terminal DC measurements at 1.6 K. Inset of (a): schematic of the electrode structure. (b) $V_{2\omega} - I_\omega$ curves for different

driving frequencies measured at 20 K. (c, d) $V_{2\omega} - B_\perp$ curves for different driving frequencies measured at 1.6 K.

## 3.Analysis of the impacts of FGT contacts on BLG

The FGT proximity can induce *n*-doping, an exchange magnetic field, and finite spin-orbit coupling (SOC) in BLG. As shown by our measurements, the additional resistance peak [Fig. 2(a)] identifies the *n*-doping on BLG. The exchange magnetic field, originating from the perpendicular magnetic anisotropy of FGT, induces nonzero spin Zeeman splitting and the valley Zeeman effect (Fig. 9). The spin Zeeman splitting lifts the spin degeneracy, resulting in spin polarization in BLG. Similarly, the orbital Zeeman effect derives from the interaction between the nonzero orbital magnetic moments of valley electrons of gapped BLG and the exchange magnetic field. It prompts nonzero valley polarization and generates currents carrying nonzero out-of-plane orbital magnetic moments. This orbital contribution reduces the observed spin torque exerted by an in-plane magnetic field compared to the previous reports. Moreover, owing to the strong SOC in FGT, electrons in the proximity-modified BLG can also experience an SOC field, further diminishing the g-factor.

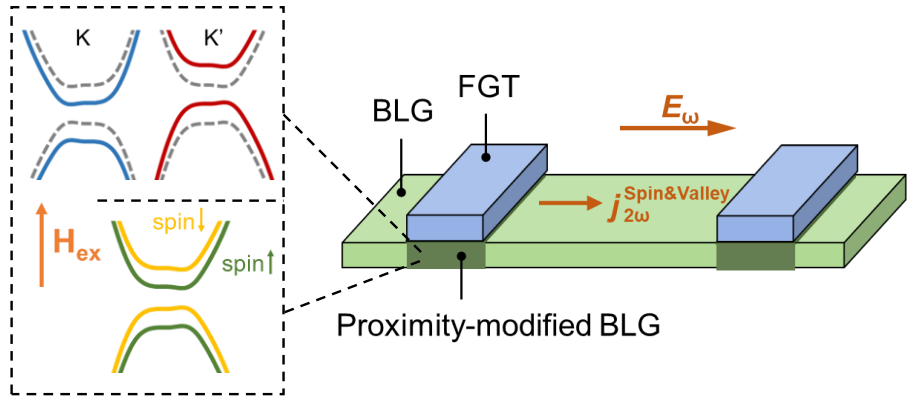

FIG. 9. Schematic diagrams of our experiments. Upon applying an applied electric field, a second-harmonic current is generated. The FGT contacts induce valley (left upper panel) and spin (left lower panel) Zeeman splitting, resulting in nonzero out-of-plane orbital and spin polarization of the second-harmonic current.

**REFERENCES**

[1] D. Xiao, W. Yao, and Q. Niu, Valley-contrasting physics in graphene: magnetic moment and topological transport, *Phys. Rev. Lett.* **99**, 236809 (2007).

[2] J. R. Schaibley, H. Yu, G. Clark, P. Rivera, J. S. Ross, K. L. Seyler, W. Yao, and X. Xu, Valleytronics in 2D materials, *Nat. Rev. Mater.* **1**, 1 (2016).

[3] L. Li, L. Shao, X. Liu, A. Gao, H. Wang, B. Zheng, G. Hou, K. Shehzad, L. Yu, and F. Miao, Room-temperature valleytronic transistor, *Nat. Nanotechnol.* **15**, 743 (2020).

[4] K. S. Novoselov, A. K. Geim, S. V. Morozov, D.-e. Jiang, Y. Zhang, S. V. Dubonos, I. V. Grigorieva, and A. A. Firsov, Electric field effect in atomically thin carbon films, *science* **306**, 666 (2004).

[5] A. H. Castro Neto, F. Guinea, N. M. Peres, K. S. Novoselov, and A. K. Geim, The electronic properties of graphene, *Rev. Mod. Phys.* **81**, 109 (2009).

[6] A. K. Geim and K. S. Novoselov, The rise of graphene, *Nat. Mater.* **6**, 183 (2007).

[7] N. Tombros, C. Jozsa, M. Popinciuc, H. T. Jonkman, and B. J. Van Wees, Electronic spin transport and spin precession in single graphene layers at room temperature, *Nature* **448**, 571 (2007).

[8] W. Han and R. K. Kawakami, Spin relaxation in single-layer and bilayer graphene, *Phys. Rev. Lett.* **107**, 047207 (2011).

[9] J. Xu, S. Singh, J. Katoch, G. Wu, T. Zhu, I. Žutić, and R. K. Kawakami, Spin inversion in graphene spin valves by gate-tunable magnetic proximity effect at one-dimensional contacts, *Nat. Commun.* **9**, 2869 (2018).

[10] W. Yao, D. Xiao, and Q. Niu, Valley-dependent optoelectronics from inversion symmetry breaking, *Phys. Rev. B* **77**, 235406 (2008).

[11] H. Zeng, J. Dai, W. Yao, D. Xiao, and X. Cui, Valley polarization in $MoS_2$ monolayers by optical pumping, *Nat. Nanotechnol.* **7**, 490 (2012).

[12] K. F. Mak, K. He, J. Shan, and T. F. Heinz, Control of valley polarization in monolayer $MoS_2$ by optical helicity, *Nat. Nanotechnol.* **7**, 494 (2012).

[13] Y. Li, J. Ludwig, T. Low, A. Chernikov, X. Cui, G. Arefe, Y. D. Kim, A. M. Van Der Zande, A. Rigosi, and H. M. Hill, Valley splitting and polarization by the Zeeman effect in monolayer $MoSe_2$, *Phys. Rev. Lett.* **113**, 266804 (2014).

[14] D. MacNeill, C. Heikes, K. F. Mak, Z. Anderson, A. Kormányos, V. Zólyomi, J. Park, and D. C. Ralph, Breaking of valley degeneracy by magnetic field in monolayer $MoSe_2$, *Phys. Rev. Lett.* **114**, 037401 (2015).

[15] G. Aivazian, Z. Gong, A. M. Jones, R.-L. Chu, J. Yan, D. G. Mandrus, C. Zhang, D. Cobden, W. Yao, and X. Xu, Magnetic control of valley pseudospin in monolayer $WSe_2$, *Nat. Phys.* **11**, 148 (2015).

[16] A. Srivastava, M. Sidler, A. V. Allain, D. S. Lembke, A. Kis, and A. Imamoğlu, Valley Zeeman effect in elementary optical excitations of monolayer $WSe_2$, *Nat. Phys.* **11**, 141 (2015).

[17] K. F. Mak, K. L. McGill, J. Park, and P. L. McEuen, The valley Hall effect in $MoS_2$ transistors, *Science* **344**, 1489 (2014).

[18] M. Sui, G. Chen, L. Ma, W.-Y. Shan, D. Tian, K. Watanabe, T. Taniguchi, X. Jin, W. Yao, and D. Xiao, Gate-tunable topological valley transport in bilayer graphene, *Nat. Phys.* **11**, 1027 (2015).

[19] Y. Shimazaki, M. Yamamoto, I. V. Borzenets, K. Watanabe, T. Taniguchi, and S. Tarucha, Generation and detection of pure valley current by electrically induced Berry curvature in bilayer graphene, *Nat. Phys.* **11**, 1032 (2015).

[20] M.-S. Qin, P.-F. Zhu, X.-G. Ye, W.-Z. Xu, Z.-H. Song, J. Liang, K. Liu, and Z.-M. Liao, Strain tunable Berry curvature dipole, orbital magnetization and nonlinear Hall effect in $WSe_2$ monolayer, *Chin. Phys. Lett.* **38**, 017301 (2021).

[21] A. Gao, Y.-F. Liu, J.-X. Qiu, B. Ghosh, T. V. Trevisan, Y. Onishi, C. Hu, T. Qian, H.-J. Tien, and S.-W. Chen, Quantum metric nonlinear Hall effect in a topological antiferromagnetic heterostructure, *Science* **381**, 181 (2023).

[22] N. Wang, D. Kaplan, Z. Zhang, T. Holder, N. Cao, A. Wang, X. Zhou, F. Zhou, Z. Jiang, and C. Zhang, Quantum-metric-induced nonlinear transport in a topological antiferromagnet, *Nature* **621**, 487 (2023).

[23] H. Isobe, S.-Y. Xu, and L. Fu, High-frequency rectification via chiral Bloch electrons, *Sci. Adv.* **6**, eaay2497 (2020).

[24] H. Yu, Y. Wu, G.-B. Liu, X. Xu, and W. Yao, Nonlinear valley and spin currents from Fermi pocket anisotropy in 2D crystals, *Phys. Rev. Lett.* **113**, 156603 (2014).

[25] J. Duan, Y. Jian, Y. Gao, H. Peng, J. Zhong, Q. Feng, J. Mao, and Y. Yao, Giant second-order nonlinear Hall effect in twisted bilayer graphene, *Phys. Rev. Lett.* **129**, 186801 (2022).

[26] P. He, G. K. W. Koon, H. Isobe, J. Y. Tan, J. Hu, A. H. C. Neto, L. Fu, and H. Yang, Graphene moiré superlattices with giant quantum nonlinearity of chiral Bloch electrons, *Nat. Nanotechnol.* **17**, 378 (2022).

[27] M. Huang, Z. Wu, X. Zhang, X. Feng, Z. Zhou, S. Wang, Y. Chen, C. Cheng, K. Sun, and Z. Y. Meng, Intrinsic nonlinear Hall effect and gate-switchable Berry curvature sliding in twisted bilayer graphene, *Phys. Rev. Lett.* **131**, 066301 (2023).

[28] I. J. Vera-Marun, V. Ranjan, and B. J. Van Wees, Nonlinear detection of spin currents in graphene with non-magnetic electrodes, *Nat. Phys.* **8**, 313 (2012).

[29] T. Ohta, A. Bostwick, T. Seyller, K. Horn, and E. Rotenberg, Controlling the electronic structure of bilayer graphene, *Science* **313**, 951 (2006).

[30] Y. Zhang, T.-T. Tang, C. Girit, Z. Hao, M. C. Martin, A. Zettl, M. F. Crommie, Y. R. Shen, and F. Wang, Direct observation of a widely tunable bandgap in bilayer graphene, *Nature* **459**, 820 (2009).

[31] E. McCann and M. Koshino, The electronic properties of bilayer graphene, *Rep. Prog. Phys.* **76**, 056503 (2013).

[32] J. Yin, C. Tan, D. Barcons-Ruiz, I. Torre, K. Watanabe, T. Taniguchi, J. C. Song, J. Hone, and F. H. Koppens, Tunable and giant valley-selective Hall effect in gapped bilayer graphene, *Science* **375**, 1398 (2022).

[33] K. Kang, T. Li, E. Sohn, J. Shan, and K. F. Mak, Nonlinear anomalous Hall effect in few-layer $WTe_2$, *Nat. Mater.* **18**, 324 (2019).

[34] Q. Ma, S.-Y. Xu, H. Shen, D. MacNeill, V. Fatemi, T.-R. Chang, A. M. Mier Valdivia, S. Wu, Z. Du, and C.-H. Hsu, Observation of the nonlinear Hall effect under time-reversal-symmetric conditions, *Nature* **565**, 337 (2019).

[35] X.-G. Ye, H. Liu, P.-F. Zhu, W.-Z. Xu, S. A. Yang, N. Shang, K. Liu, and Z.-M. Liao, Control over Berry curvature dipole with electric field in $WTe_2$, *Phys. Rev. Lett.* **130**, 016301 (2023).

[36] D. Kumar, C.-H. Hsu, R. Sharma, T.-R. Chang, P. Yu, J. Wang, G. Eda, G. Liang, and H. Yang, Room-temperature nonlinear Hall effect and wireless radiofrequency rectification in Weyl semimetal $TaIrTe_4$, *Nat. Nanotechnol.* **16**, 421 (2021).

[37] S. Das Sarma, S. Adam, E. Hwang, and E. Rossi, Electronic transport in two-dimensional graphene, *Rev. Mod. Phys.* **83**, 407 (2011).

[38] Z. Du, C. Wang, S. Li, H.-Z. Lu, and X. Xie, Disorder-induced nonlinear Hall effect with time-reversal symmetry, *Nat. Commun.* **10**, 3047 (2019).

[39] R. Gorbachev, F. Tikhonenko, A. Mayorov, D. Horsell, and A. Savchenko, Weak localisation in bilayer graphene, *Phys. E: Low-Dimens. Syst. Nanostructures* **40**, 1360 (2008).

[40] Z. Fei, B. Huang, P. Malinowski, W. Wang, T. Song, J. Sanchez, W. Yao, D. Xiao, X. Zhu, and A. F. May, Two-dimensional itinerant ferromagnetism in atomically thin $Fe_3GeTe_2$, *Nat. Mater.* **17**, 778 (2018).

[41] Y. Deng, Y. Yu, Y. Song, J. Zhang, N. Z. Wang, Z. Sun, Y. Yi, Y. Z. Wu, S. Wu, and J. Zhu, Gate-tunable room-temperature ferromagnetism in two-dimensional $Fe_3GeTe_2$, *Nature* **563**, 94 (2018).

[42] C. Tan, J. Lee, S.-G. Jung, T. Park, S. Albarakati, J. Partridge, M. R. Field, D. G. McCulloch, L. Wang, and C. Lee, Hard magnetic properties in nanoflake van der Waals $Fe_3GeTe_2$, *Nat. Commun.* **9**, 1554 (2018).

[43] X. He, C. Zhang, D. Zheng, P. Li, J. Q. Xiao, and X. Zhang, Nonlocal Spin Valves Based on Graphene/Fe$_3$GeTe$_2$ van der Waals Heterostructures, *ACS Appl. Mater. Interfaces* **15**, 9649 (2023).

[44] H. Pan, C. Zhang, J. Shi, X. Hu, N. Wang, L. An, R. Duan, P. Deb, Z. Liu, and W. Gao, Room-temperature lateral spin valve in graphene/Fe$_3$GaTe$_2$ van der Waals heterostructures, *ACS Mater. Lett.* **5**, 2226 (2023).

[45] F. Jedema, H. Heersche, A. Filip, J. Baselmans, and B. Van Wees, Electrical detection of spin precession in a metallic mesoscopic spin valve, *Nature* **416**, 713 (2002).

[46] T.-Y. Yang, J. Balakrishnan, F. Volmer, A. Avsar, M. Jaiswal, J. Samm, S. Ali, A. Pachoud, M. Zeng, and M. Popinciuc, Observation of long spin-relaxation times in bilayer graphene at room temperature, *Phys. Rev. Lett.* **107**, 047206 (2011).

[47] P. Wei, S. Lee, F. Lemaitre, L. Pinel, D. Cutaia, W. Cha, F. Katmis, Y. Zhu, D. Heiman, and J. Hone, Strong interfacial exchange field in the graphene/EuS heterostructure, *Nat. Mater.* **15**, 711 (2016).

[48] Y. Zhang, Y.-W. Tan, H. L. Stormer, and P. Kim, Experimental observation of the quantum Hall effect and Berry's phase in graphene, *Nature* **438**, 201 (2005).

[49] K. S. Novoselov, E. McCann, S. Morozov, V. I. Fal'ko, M. Katsnelson, U. Zeitler, D. Jiang, F. Schedin, and A. Geim, Unconventional quantum Hall effect and Berry's phase of $2\pi$ in bilayer graphene, *Nat. Phys.* **2**, 177 (2006).

[50] M. Jonson and G. Mahan, Mott's formula for the thermopower and the Wiedemann-Franz law, *Phys. Rev. B* **21**, 4223 (1980).